\documentclass{article}
\usepackage{spconf,amsmath,graphicx}
\usepackage{adjustbox}

\usepackage{svg}
\usepackage{amsmath,amssymb,amsfonts}

\usepackage{tikz}
\usepackage{cite}

\usepackage {cancel} 

\usepackage{xcolor}
\usepackage{hyperref}
\usepackage[utf8]{inputenc}
\usepackage{pgfplots}
\usepackage{caption}
\usepackage{subcaption}
\usepackage{amsthm}
\usepackage[inline]{enumitem}

\DeclareUnicodeCharacter{2212}{−}
\usepgfplotslibrary{groupplots,dateplot}
\usetikzlibrary{patterns,shapes.arrows}
\pgfplotsset{compat=newest}

\definecolor{brown1926061}{RGB}{192,60,61}
\definecolor{darkgray176}{RGB}{176,176,176}
\definecolor{darkslategray61}{RGB}{61,61,61}
\definecolor{lightgray204}{RGB}{204,204,204}
\definecolor{peru22412844}{RGB}{224,128,44}
\definecolor{seagreen5814558}{RGB}{58,145,58}
\definecolor{steelblue49115161}{RGB}{49,115,161}
    \newcommand{\cstN}{N}

    \newcommand{\varI}{i}
    \newcommand{\varJ}{j}
    \newcommand{\varK}{k}

    \newcommand{\graphG}{{\mathcal G}}
    \newcommand{\setV}{{\mathcal V}}
    \newcommand{\setE}{{\mathcal E}}

    \newcommand{\vecSs}{{\bf s}}
    \newcommand{\vecs}{{s}}
    
     \newcommand{\vecSshat}{{\bf \hat{s}}}

    \newcommand{\matW}{{\bf W}}
    \newcommand{\matD}{{\bf D}}

    \newcommand{\matL}{{\bf L}}
    \newcommand{\matU}{{\bf U}}
    \newcommand{\Epsi}{{\mathcal{M}}}
    \newcommand{\bEpsi}{{\bcancel{\mathcal{M}}}}

\newtheorem{proposition}{Proposition}

    \newcommand{\matLambda}{{\boldsymbol \Lambda}}

        \newcommand{\R}{{\mathbb R}}

    \newcommand{\set}[1]{\left\{ #1 \right\}}
    \newcommand{\tuple}[1]{\left\langle #1 \right\rangle}
    \newcommand{\card}[1]{\left| #1 \right|}

        \DeclareMathOperator{\gft}{GFT}

\let\originalleft\left
\let\originalright\right
\renewcommand{\left}{\mathopen{}\mathclose\bgroup\originalleft}
\renewcommand{\right}{\aftergroup\egroup\originalright}

\title{Spatial Graph Signal Interpolation with an Application for Merging BCI Datasets with Various Dimensionalities}
\name{Yassine El Ouahidi, Lucas Drumetz, Giulia Lioi, Nicolas Farrugia, Bastien Pasdeloup and Vincent Gripon \thanks{Thanks to the Brittany region for its support.}}

\address{IMT Atlantique, Lab-STICC, UMR CNRS 6285, F-29238 Brest, France, name.surname@imt-atlantique.fr}
\begin{document}
%
\maketitle
\begin{abstract}
  BCI Motor Imagery datasets usually are small and have different electrodes setups. When training a Deep Neural Network, one may want to capitalize on all these datasets to increase the amount of data available and hence obtain good generalization results.
  To this end, we introduce a spatial graph signal interpolation technique, that allows to interpolate efficiently multiple electrodes. We conduct a set of experiments with five BCI Motor Imagery datasets comparing the proposed interpolation with spherical splines interpolation. We believe that this work provides novel ideas on how to leverage graphs to interpolate electrodes and on how to homogenize multiple datasets.

\end{abstract}
\section{Introduction}
\label{sec:intro}

One important challenge in brain signals classification is the lack of large, homogeneous datasets, such as those available in computer vision. This is also true for Brain Computer Interfaces (BCI) where usually the electroencephalographic (EEG) signal is collected from a few subjects with specific recording setups (electrodes layout, sampling frequency, stimulation paradigm). The lack of training data also explain why only recently methods based on deep learning and transfer learning have been applied to BCI signal decoding, with contrasting results~\cite{lotte2018review}. One way to tackle this limitation is to setup a joint analysis on datasets. Such solutions usually rely on handcrafted features based on physiological priors or hardly interpretable deep learning models. In this paper, we propose a new method for merging EEG datasets based on graph signal interpolation and show an application to BCI Motor Imagery (MI) classification. A graph is learned from EEG data and a graph signal interpolation technique is implemented to obtain a unified, virtual, electrodes setup. This method does not require any prior and provides interpretable results in terms of EEG spatial patterns.

\section{Related work}

\subsection{BCI MI classification}
\label{sec:related_BCI}
BCI systems are based on the real-time measure of brain signals (typically EEG) and usually need an offline training phase during which the system is calibrated; then during the operational phase the system classifies brain activity patterns and translates them into commands for a device in real-time. In current applications, a training dataset needs to be pre-recorded from the user to have reliable systems, due to the lack of classifiers able to generalise across subjects and EEG setups~\cite{lotte2018review}. To go towards calibration-free, high accuracy BCI systems, a key feature is the ability to efficiently extract knowledge from the variety of datasets available in literature and transfer it to new subjects. However, the great majority of the works in literature consider separate BCI datasets~\cite{Kostas_2020,Lawhern_2018}. 

Recently, an attempt to exploit information jointly from different EEG datasets has been made with the NeurIPS EEG Transfer learning challenge BEETL~\cite{Wei20212021BC}. This challenge aims at developing algorithms to transfer knowledge between different subjects and EEG datasets and at defining a new benchmark for EEG signals classification. In this challenge, three source MI datasets are provided, and algorithms are then tested on two unseen datasets.  In order to train a unique model on the ensemble of the source MI datasets, a first step is to integrate data from different setups. In the context of the challenge, simple solutions such as considering common electrodes (intersection) were proposed, which totally disregards information of the dropped electrodes. Here we propose a methodology that exploits information from all the available electrodes based on graph signal interpolation.

\subsection{Graph Signal Processing}

An intuitive way to represent interactions between electrodes in BCI context is to use spatial graphs. EEG signals can then be seen as observations over this graph, with an added temporal dimension. Graph Signal Processing (GSP) then offers the tools to process such signals~\cite{shuman2013emerging, ortega2018graph}. In the context of MI decoding, the use of GSP-based methods also brings forward interpretability questions.
In this setting, a weighted graph $\graphG = \tuple{\setV, \setE}$ with vertices $\setV $ and edges $\setE \subset \setV \times \setV$ is used to model electrodes and their interactions. Such a graph can be equivalently represented by a symmetric weights matrix $\matW \in \R^{\card\setV \times \card\setV}$ such that $W_{ij} = 0$ if $\set{i,j} \not\in \setE$. We then note $\matD \in \R^{\card\setV \times \card\setV}$ the degrees matrix of $\graphG$, such that $D_{ij} = \sum_{\varK = 1}^{\card\setV} W_{ik}$ if $\varI = \varJ$ and $0$ otherwise.
From these two matrices, we can compute the Laplacian $\matL = \matD - \matW$ of $\graphG$. Since $\matL$ is real and symmetric, it can be diagonalized as $\matL = \matU \matLambda \matU^\top$, where $\matU$ is a matrix of orthonormal vectors associated with eigenvalues forming the diagonal matrix $\matLambda$, sorted in increasing order.
A signal $\vecSs \in \R^\cstN$ on $\graphG$ is an observation on each of its vertices.  Its Graph Fourier Transform $\vecSshat = \gft(\vecSs) = \matU^\top \vecSs$ can be seen as an observation for each graph frequency.
Using those elements we can define the total graph signal variation $\sigma(\vecSs)$ of a signal $\vecSs$ as:
\begin{equation}
        \vspace{-3mm} 
\sigma(\vecSs)= \vecSs^\top \matL \vecSs = \sum_{i=1}^{\card\setV}\Lambda_{ii} \hat{s}_{i}^{2} = \sum_{i=1}^{\card\setV}\sum_{j=1}^{\card\setV}W_{ij}(s_{i} - s_{j})^{2}
\label{eq:variation}
\end{equation}


%
\subsection{Interpolation of EEG Signals}

Interpolating electrodes is a common step in many EEG preprocessing pipelines and is usually needed to artificially repair the signal from noisy electrodes or trials. Multiple methods allow to perform spatial interpolation without using graphs. The reference method, implemented in many pipelines, is the Spherical Spline interpolation~\cite{perrin1989spherical} with Legendre polynomials. Other methods use signal correlation~\cite{ramakrishnan2016reconstruction} or deep learning, like~\cite{corley2018deep} which is using a deep learning generator, or~\cite{kwon2019super} with a deep convolutional network. However, these methods suffer from either high reconstruction error or lack of interpretability. Other methods allow to perform interpolation with the help of graphs, like the work of~\cite{tang2022deep}, where structural and functional connectivities are used along with a deep graph convolutional network. Their method is designed to interpolate in multiple dimensions (frequency, spatial and temporal).  Other works try to optimize a graph signal criterion, like sparsity ~\cite{humbert2019subsampling}, or a non smooth criterion \cite{mazarguil2022non}, but the last method is designed to perform temporal interpolation.


\subsection{Homogenizing BCI datasets}
\label{sec:homog} 
Few approaches have been explored to solve the problem introduced in Section~\ref{sec:related_BCI} of exploiting information jointly from different EEG datasets. In general, datasets differ in several ways (e.g  sampling frequency, preprocessing, recording paradigm, electrode setup). 
Some of these inhomogeneities can be solved by simply windowing or resampling the EEG signals, while unifying the electrode setup remains a challenge. To solve it, different methods have been proposed. The BEETL challenge~\cite{wei20222021} winners simply reduce the spatial dimension by taking the intersections of the datasets. ~\cite{nguyen2022combining} performs PCA, which also implies a loss of information. To overcome this limitation, some techniques propose to learn another representation with adversarial learning~\cite{bethge2022exploiting} while others propose to augment the spatial dimension by going into the Riemannian space~\cite{bethge2022exploiting}.
We propose to increase the spatial dimension by staying in the electrode space and interpolate the electrodes using a graph variation optimization.

\section{Methodology}

Our approach consists in considering each BCI dataset as a partial sampling of a virtual, larger collection of electrodes and then  using a graph interpolation technique to ``recover'' the missing electrodes.
The graph is learned from a different EEG dataset containing all electrodes. We then use this graph to homogeneize the considered BCI datasets and train them altogether in a similar setup as for the BEETL challenge.

\subsection{Graph Signal Interpolation}

We propose a graph signal interpolation technique that consists of minimizing graph signal variation in Equation~\eqref{eq:variation} while only knowing a part of the signal. We show here that this problem admits a closed form.

Consider a graph $\graphG$ where electrodes are vertices $\setV$, with Laplacian matrix $\matL\in \R^{\card\setV \times \card\setV}$. Let $\vecSs\in \R^{\card\setV}$ be a signal on $\graphG$. In our problem, we consider that $\vecSs$ has some missing entries, due to absence of some electrodes. The set of such missing electrodes is denoted $\Epsi \subset \setV$, and its complement $\bEpsi =  \setV \setminus \Epsi$.
We note $\vecSs_{\bEpsi}= \set{\vecs_{i\in \bEpsi}}$  the observed part of $\vecSs$, and $\vecSs_{\Epsi}= \set{\vecs_{i\in\Epsi}}$ its missing part. 
We note $\matL_{\Epsi}\in\R^{\card \Epsi \times \card\Epsi}$ the submatrix of $\matL$
where we only keep the rows and columns with indices in $\Epsi$, and $\matL_{\Epsi\bEpsi}\in\R^{\card\Epsi \times \card\bEpsi}$ the submatrix of $\matL$ 
where we only keep the rows with indices in $\Epsi$, and the columns with indices in $\bEpsi$.

\begin{proposition}
The solution $\vecSs_{\Epsi}$ that optimizes the variation problem Equation~\eqref{eq:variation} in our setup is directly given by the following closed form:
\begin{equation}
    \vecSs_{\Epsi}=- \matL_{\Epsi}^{-1} \matL_{\Epsi\bEpsi}\vecSs_{\bEpsi}
\label{eq:closedform}
\end{equation}
\end{proposition}

\begin{proof}

We start from:
\begin{equation}
\sigma(\vecSs)= \vecSs^\top \matL \vecSs=\sum_{i=1}^{\card\setV}\sum_{j=1}^{\card\setV}\vecs_{i}\matL_{ij}\vecs_{j}
\label{eq:starteq}
\end{equation}
We can decompose Equation~\eqref{eq:starteq} into 4  terms: \begin{enumerate*} \item $i\in\Epsi$ and $j\in\Epsi$; \item $i\in\Epsi$ and $j\notin\Epsi$; \item $i\notin\Epsi$ and $j\in\Epsi$; \item $i\notin\Epsi$ and $j\notin\Epsi$. \end{enumerate*} In our setup 2. and 3. are symmetric, and 4. is a constant.
So we have: 
\begin{equation}
  \sigma(\vecSs)= \vecSs_\Epsi^\top \matL_\Epsi \vecSs_\Epsi + 2 \sum_{i\in\Epsi}\sum_{j\notin\Epsi}\vecs_{i}\matL_{ij}\vecs_{j} + constant
  \;,
\label{eq:dvpt}
\end{equation}
where $\sigma_{\Epsi \bEpsi}(\mathbf{s})=2\sum_{i\in\Epsi}\sum_{j\notin\Epsi}\vecs_{i}\matL_{i j}\vecs_{j}$  corresponds to the two symmetric parts 2. and 3. in Equation~\eqref{eq:dvpt}. We have: 

\begin{equation} 
\forall i \in \Epsi, \frac{\partial \sigma_{\Epsi \bEpsi}}{\partial \vecSs_{i}}=2\sum_{j\notin\Epsi}L_{i j}\vecSs_j=2\left(\matL_{\Epsi\bEpsi}\vecs_{\bEpsi}\right)_i
\;,
\label{eq:part23}
\end{equation}
where we denote as $(\cdot)_i$ the $i^\textrm{th}$ entry of the vector in parentheses. We then obtain:  

\vspace{-.5cm}
\begin{equation}
\small
\nabla\sigma(\vecSs_{\Epsi}) = 2(\matL_{\Epsi}\vecSs_{\Epsi}+\matL_{\Epsi\bEpsi}\vecSs_{\bEpsi})=0 \Rightarrow \vecSs_{\Epsi}= - \matL_{\Epsi}^{-1} \matL_{\Epsi\bEpsi}\vecSs_{\bEpsi}
\label{eq:end}
\end{equation}

\end{proof}

\vspace{-.5cm}
Reconstructing entries in $\Epsi$ therefore requires the use of a graph. To do so, we propose to learn it from EEG signals.

\subsection{Learning the graph}

 In order to reconstruct the missing entries $\Epsi$ in a signal, we propose to learn a graph $\graphG$ of $\setV$ electrodes. We consider here three different graphs:
 
 \begin{enumerate}
 
 \item A spatial graph, where the adjacency matrix is based on the locations of the sensors. We binarize this graph based on a radius. This graph has the same benefits as the method Spherical Spline interpolation, and does not need additional data except the location of the sensors;
 
 \item We add information on top of the spatial graph, by weighting its edges with Weighted Phase Lag Index (WPLI) \cite{wpli} scores computed from a dataset featuring all the electrodes. A similar graph was proposed in \cite{menoret2017evaluating};
 
 \item We learn the adjacency matrix of an optimal graph by gradient descent, in order to build a graph that optimizes our signal variation problem. We use as loss function $1-R^2$, with $R^2$ the coefficient of determination. At each step of the learning process, we randomly mask half of the electrodes and try to reconstruct them. 
\end{enumerate}

In a first step, we propose to learn the graph using a large dataset A containing all electrodes. In a second step, we use the learned graph to reconstruct missing electrodes in another dataset B. For that we experiment two ways of doing it. \begin{enumerate*} \item Directly transferring, which means using the learned graph on A to reconstruct B; \item Fine-tuning the learned graph A towards the dataset B. For this, a second phase where we jointly train using A and B is needed to adapt the graph on B. At each step, we solve two different problems. The first one is on B where, we randomly mask half of the electrodes of the dataset B and we try to reconstruct them. The second problem is on A where we mask all the electrodes which are A but not in B, and we try to reconstruct them. The loss is therefore a weighted average of the losses of the two problems (each loss is still $1-R^2$). Doing that helps prevent overfitting on dataset B while fine tuning. It also helps to align dataset B to dataset A which will help to homogenize multiple datasets in the second set of experiments. \end{enumerate*}

Usefulness of the graphs for reconstructing missing entries is evaluated on controlled problems, where we artificially mask electrodes, and evaluate the reconstruction capabilities of a given graph.

\section{Experiments}

\subsection{Interpolation of electrodes}
\label{sec:exp-interpo}

We used multiple open access EEG MI datasets. Which are all available on MOABB \cite{jayaram2018moabb} \footnote{\url{http://moabb.neurotechx.com}}. The code of our experiments is also available at our Github\footnote{\url{https://github.com/elouayas/eeg_interpolation}}.

\begin{table}[!h]
\scalebox{0.66}{
\centering
\begin{tabular}{lllllll}
\hline
MI dataset & Subjects & Electrodes & Samples & \begin{tabular}[c]{@{}l@{}}Sampling\\ frequency\end{tabular} & Dur. & Tasks      \\ \hline
Schrimister~\cite{schirrmeister2017deep} & 14       & 76       & 13484   & 500 Hz                                                       & 4 s.     & L/R/F/R    \\
BNCI2014~\cite{tangermann2012review}    & 9        & 22       & 5184    & 160 Hz                                                       & 3 s.     & L/R/F/T    \\
Cho2017~\cite{cho2017eeg}     & 52       & 64       & 9880    & 160 Hz                                                       & 3 s.     & L/R        \\
PhysionetMI~\cite{goldberger2000physiobank} & 109      & 62       & 9838    & 160 Hz                                                       & 3 s.     & L/R/F/BH/R \\
Zhou2016~\cite{zhou2016fully}    & 4        & 14       & 1800    & 160 Hz                                                       & 3 s.     & L/R/F      \\
Shin2017~\cite{shin2016open}    & 29       & 22       & 1740    & 160 Hz                                                       & 3 s.     & L/R        \\ \hline
\end{tabular}}
\vspace{-4mm} 
 \caption{MI datasets considered. L = Left hand, R = Right hand, F = Feet, R = Rest, BH = Both hands}
\end{table}

We have chosen to use the Schrimister dataset to learn the union graph. This dataset has the particularity to contain almost all of the electrodes of the datasets that we consider later in our application case. This dataset is also used to learn the optimal graph, and to compute WPLI. We take all the data available and apply a band-pass filter at [2,40] Hz. All the other datasets are used for two purposes: \begin{enumerate*} \item To evaluate the performance of various interpolation techniques Section~\ref{sec:exp-interpo}; \item To create a similar setup than in the BEETL challenge and evaluate the performance of our interpolation method to homogenize datasets Section~\ref{sec:exp-multi-training}. \end{enumerate*} In order to homogenize the 5 remaining datasets, we downsample to the lowest sampling frequency (160 Hz), use the first 3 seconds starting at the cue, and apply a [2,40] Hz band-pass filter. 

\begin{figure*}[!h]
     \centering
        \begin{subfigure}[b]{0.33\textwidth}
            \centering
            \scalebox{0.7}{\input{images/r2.tex}}
            \label{fig:r2}
            \caption{}
        \end{subfigure}
        \begin{subfigure}[b]{0.33\textwidth}
            \centering
             \scalebox{0.7}{\input{images/mse.tex}}
            \label{fig:mse}
            \caption{}
        \end{subfigure}
        \begin{subfigure}[b]{0.33\textwidth}
            \centering
            \includegraphics[width=0.8\linewidth]{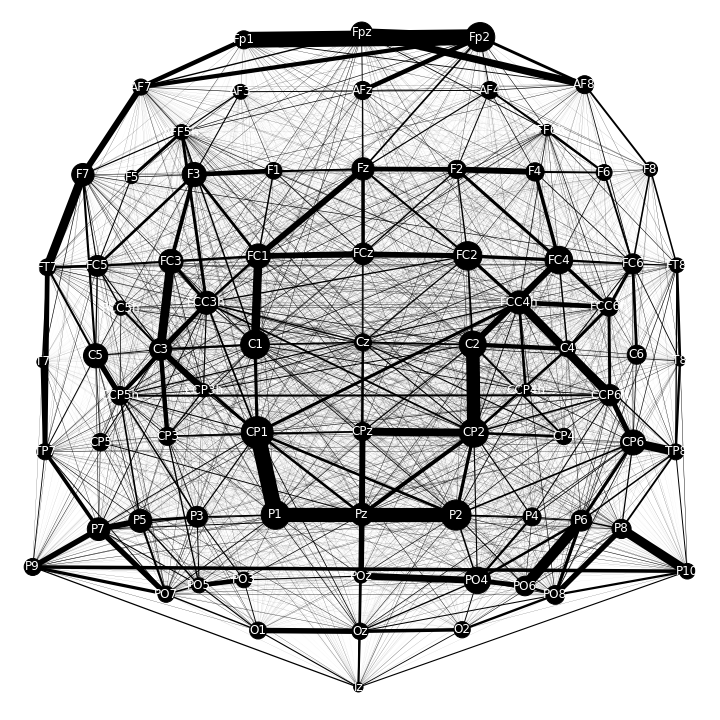}
            \label{fig:graph}
            \caption{}
        \end{subfigure}
        \caption{(a, b) Reconstruction performances for various number of missing electrodes on Schrimister (among 76 electrodes). (c) Learned graph. The size of the nodes is weighted by their strengh, and the thickness correspond to their weights. Both of them are in a square scale in order to accentuate the gaps.}
    \label{fig:perfs}
\end{figure*}

We notice in Fig.~\ref{fig:perfs} that the WPLI weighted graph allows to slightly better reconstruct the signal than the spatial graph. On the other hand, the Spherical Spline method is more efficient than the spatial and WPLI graph. We show that learning the graph outperforms all other graphs even with up to 57 missing electrodes. In the rest of this paper, we will therefore consider the learned graph.

We then evaluate the ability  of the learned graph to reconstruct signals on the other datasets of our study. In Table~\ref{table:transfer}, we observe that directly using it (the Transfer line), is not very efficient. This indicates that the graph is over-adapted to the Schrimister dataset on which it is trained. To compensate for this, we fine-tune it as described in Section~\ref{sec:homog} to each dataset individually.

By doing so, we prove the interest of the method that consistently outperforms (the Transfer + FT line) the Spherical Spline method, on all datasets, except BNCI where the fine tuning results are similar.
In Fig.~\ref{fig:perfs}, we display the learned graph after it is trained on the Schrimister dataset. We observe a cluster in the Centro-Fronto-Parietal area on the right side around C4, and symmetrically on the left side around C3. Those observations make sense as those are regions and connections recruited during a motor imagery task ~\cite{cury2020impact}.

\begin{table}[!h]
\centering
    \scalebox{0.7}{
\begin{tabular}{l|l|l|l|l|l|}
\cline{2-6}
                                            & Physionet~\cite{goldberger2000physiobank} & BNCI~\cite{tangermann2012review} & Cho~\cite{cho2017eeg} & Shin~\cite{shin2016open} & Zhou~\cite{zhou2016fully}   \\ \hline
\multicolumn{1}{|l|}{Transfer}      & 52.5$\pm$3.6      & 56.3$\pm$30.4     & 18.8$\pm$11.8     & 33.8$\pm$17.3      & 8.3$\pm$18.1              \\ \hline
\multicolumn{1}{|l|}{Transfer + FT} & \bf{86.1$\pm$1.2} & 93.2$\pm$2.8      & \bf{77.1$\pm$2.3} & \bf{90.4$\pm$1.7}  & \bf{70.0$\pm$9.2}             \\ \hline
\multicolumn{1}{|l|}{Spatial}        & 59.3$\pm$2.6      & 79.3$\pm$4.8      & 35.0$\pm$6.8      & 73.4$\pm$4.2       & 32.9$\pm$18.4              \\ \hline
\multicolumn{1}{|l|}{Spherical}     & 80.6$\pm$4.5      & \bf{94.9$\pm$1.5} & 75.6$\pm$4.4      & 85.9$\pm$4.2       & 54.1$\pm$21.8             \\ \hline
\end{tabular}}
\vspace{-2mm} 
 \caption{Reconstruction $R^2$ of half of the electrodes for various methods and various datasets}
 \label{table:transfer}
 \vspace{-.5cm}
\end{table}




\subsection{Application to multi BCI dataset training}
\label{sec:exp-multi-training}


In order to evaluate the efficiency of our interpolation method to homogenize different BCI datasets, we decided to use a setup similar to the BEETL challenge. For this, we took the same 3 source datasets (BNCI, Cho and PhysionetMI) and two new target datasets (Shin and Zhou). We use only the first 40 trials for each subjects of the two target datasets, and try to generalize to the rest of the dataset. 
We propose to use a deep learning model to classify our data. For this we train a 1D CNN jointly on the 4 datasets. The input of the model has for dimension the number of electrodes, so the signals of each electrode enter individually in a 1D convolution channel. The network is composed of two parts: \begin{enumerate*} \item A backbone common to all datasets (with a unique entry, and multiple convolutional layers); \item Multiple classification heads unique to each dataset (3 for the source datasets and 1 for the target dataset). \end{enumerate*}

We compare through this setup the classification performances of taking only the intersection of electrodes of the 4 datasets to the performances of interpolating the missing electrodes at the intersection. We compare two way of interpolating, the first one consists in interpolating all the electrodes present in the 4 datasets and the second one consists in interpolating and selecting only the electrodes of the target dataset which is evaluated.
Before training our models we align our data with Euclidean alignment~\cite{he2019transfer}, and resample the 4 datasets by oversampling the smaller datasets.


In Table~\ref{table:perf_intersectionunion}, we notice that interpolating only the electrodes of the target dataset is more efficient than interpolating all the electrodes of the 4 datasets. This could be due to the fact that the information needed to classify a target dataset is contained within its electrodes, and that it is not necessary to go and get it artificially in those of the others. On the other hand it is useful to be able to exploit at least all its electrodes by interpolating the 3 source datasets.
Another conclusion is that it is not always worth interpolating using our method and this procedure. If we look at the Zhou results, we see that the intersection is the one which is performing the better. This difference in results could be explained by the fact that the difference between electrodes intersection and union is very drastic for Shin, differently from Zhou.

\begin{table}[!h]
\vspace{-.2cm}
\centering
\begin{tabular}{c|cc|cc|}
\cline{2-5}
\multicolumn{1}{l|}{}              & \multicolumn{2}{c|}{Shin~\cite{shin2016open}}                                    & \multicolumn{2}{c|}{Zhou~\cite{zhou2016fully}}                                    \\ \cline{2-5} 
\multicolumn{1}{l|}{}              & \multicolumn{1}{c|}{Accuracy}                           & N  & \multicolumn{1}{c|}{Accuracy}                           & N  \\ \hline
\multicolumn{1}{|c|}{Intersection} & \multicolumn{1}{c|}{53.2$\pm$2.8}                       & 2  & \multicolumn{1}{c|}{\bf{61.2$\pm$2.0}} & 9  \\ \hline
\multicolumn{1}{|c|}{Dataset}  & \multicolumn{1}{c|}{\textbf{63.2$\pm$2.3}} & 22 & \multicolumn{1}{c|}{56.2$\pm$4.8}                       & 14 \\ \hline
\multicolumn{1}{|c|}{Union}    & \multicolumn{1}{c|}{62.3$\pm$2.1}                       & 76 & \multicolumn{1}{c|}{46.4$\pm$2.8}                       & 76 \\ \hline
\end{tabular}
\vspace{-.3cm} 
 \caption{Generalization accuracy on the two target dataset. N stands for number of electrodes. "Intersection" stands for the electrodes common to the 4 datasets. "Dataset" corresponds to the electrodes of the dataset in the column. "Union" is the union of electrodes in all 5 datasets.}
 \label{table:perf_intersectionunion}
\end{table}

\section{Conclusion}

A new and efficient electrode interpolation technique exploiting GSP tools has been proposed. We have illustrated the interest of our method to homogenize datasets. Our method allows to interpolate electrodes efficiently, and some results open new questions, especially on how one should homogenize datasets in the case where the intersection of electrodes is not reduced to its bare minimum.

\vfill\pagebreak


\footnotesize
\bibliographystyle{IEEEbib}
\bibliography{refs}

\end{document}